\documentclass[useAMS,usenatbib]{mn2e}
\usepackage{epsf}
\usepackage{amssymb}
\usepackage{epsfig}
\usepackage[usenames]{color}
\usepackage{float}
\title[{\it Chandra}, {\it XMM-Newton} and {\it RXTE} observations of 4U 1626-67]{Luminosity dependent change of the emission diagram in the X-ray pulsar 4U 1626-67}
\author[F.Koliopanos, M.Gilfanov]{Filippos Koliopanos$^{1,2,3}$\thanks{filippos@mpa-garching.mpg.de} and Marat Gilfanov$^{1,4,5}$ \\
$^{1}$MPI f\"ur Astrophysik, Karl-Schwarzschild str. 1, Garching, 85741, Germany\\
$^{2}$Universit{\'e} de Toulouse; UPS-OMP; IRAP, Toulouse, France\\
$^{3}$CNRS, IRAP, 9 Av. colonel Roche, BP 44346, F-31028 Toulouse cedex 4, France\\
$^{4}$Space Research Institute of Russian Academy of Sciences, Profsoyuznaya 84/32, 117997 Moscow, Russia\\
$^{5}$Kazan Federal University, Kremlevskaya str.18, 420008 Kazan, Russia\\  }

\begin{document}

\date{Accepted .... Received ...}

\pagerange{\pageref{firstpage}--\pageref{lastpage}} \pubyear{2015}

\maketitle

\label{firstpage}

\begin{abstract}
We detect variability of the Fe ${\rm K\alpha}$ emission line in the spectrum of X-ray pulsar 4U 1626-67,
correlated with  changes in its luminosity and  in the shape of its pulse profile. 
Analysis of archival {\it Chandra} and {\it RXTE} observations revealed the presence of an intrinsically narrow  Fe ${\rm K\alpha}$ emission line in the spectrum obtained during the source's current high luminosity period. 
However, the line was not present during an {\it XMM-Newton} observation seven years earlier, when the source  was $\sim$three times fainter. The line is resolved by  the high energy grating of {\it Chandra} at the 98\%  confidence level, and its small intrinsic width, $\sigma=36.4_{-11.3}^{+15.3}$\,eV, suggests reflection off an accretion disk at the radius $R\approx (7.5_{-3.8}^{+8.2})\times 10^8$\,cm assuming a Keplerian disk,  viewed at an inclination angle of $20\degr$. This value is consistent with the radius of the  magnetosphere of the pulsar, suggesting that the line originates near 
the inner edge of a disk that is truncated by the magnetic field of the neutron star.
Timing analysis of the {\it XMM-Newton} and {\it RXTE} data revealed 
a major change in the pulse profile of the source from a distinct   double peaked shape during the high luminosity state when the line was present,  to a much more complex multi-peak structure  during the low luminosity state.
We argue that the appearance of the line and the change in the shape of the pulse profile are correlated
and are the result of a major change in the emission diagram of the accretion column, from a pencil-beam pattern at low luminosity, to a fan-beam  pattern at high luminosity.

\end{abstract}

\begin{keywords}
Keywords from the MNRAS website
\end{keywords}

\section{Introduction}
4U 1626-67 is an accreting X-ray pulsar located at a distance of $\sim5-13$\,kpc from the Sun \citep{1998ApJ...492..342C}.
It has a pulsation period of $P_{\rm spin}\approx$7.7\,s and an orbital period of
$P_{\rm orb}\approx$42\, min, \citep[][]{1981ApJ...244.1001M, 1998ApJ...492..342C}.   
Its short orbital period classifies it as an ultra compact X-ray binary (UCXB), a subgroup of low mass X-ray binaries (LMXBs) with orbital periods of less than one hour.
The short orbital periods of UCXBs suggest such tight orbits that only an evolved compact donor can fit. UCXBs, most likely,
consist of a white dwarf or a helium star that is accreting onto a neutron star due to Roche lobe overflow 
\citep*[e.g.][]{1993ARep...37..411T, 1995ApJS..100..233I, 1995xrbi.nasa..457V, 2003ApJ...598.1217D, 2005ApJ...624..934D}.
Due to their evolved nature, UCXB donors are expected to be hydrogen deficient. Depending on the initial conditions and the environment in which they are created 
(e.g. being part of a globular cluster) their donors can follow different evolutionary paths, leading to a variety of objects, ranging from non-degenerate He stars
to C/O or O/Ne/Mg white dwarfs \citep*[e.g.][]{1986A&A...155...51S,2002ApJ...565.1107P,2002A&A...388..546Y, 2004ApJ...607L.119B}.

Since its discovery \citep{1972ApJ...178..281G, 1977ApJ...217L..29R}, 4U 1626-67 has been observed by all major X-ray observatories. Furthermore, UV and optical spectra have been obtained
by the {\it Hubble Space Telescope} and the {\it Very Large Telescope}, respectively. Analysis of BeppoSAX observations by \cite{1998ApJ...500L.163O} revealed the presence of
cyclotron absorption lines, suggesting  strong magnetic field of $\approx3\times10^{12}$\,G, making the source  the only  strongly magnetised neutron star in an ultracompact binary known  so far. 
Another intriguing feature of its X-ray spectrum  is the presence of prominent Ne and O emission lines. First detected in ASCA data, analyzed by 
\cite{1995ApJ...449L..41A}, they have been studied extensively using high resolution spectroscopy \citep{2001ApJ...563..941S, 2007ApJ...660..605K}. 
The lines are believed to originate in C/O or O/Ne dominated hot plasma in the vicinity of the  compact object, 
and suggest a donor whose chemical composition is enriched by products of later stages of nuclear burning,
most likely, a C-O-Ne or O-Ne-Mg white dwarf. These findings are further corroborated by the HST UV spectrum that revealed 
both emission and absorption features from C, O, and Si but lacked He emission lines \citep{2002AJ....124.3348H} and
by the VLT optical spectrum that featured prominent C and O emission lines, but showed no evidence of H or He \citep{2006A&A...450..725W}.

The timing properties and the shape of the pulse profile of 4U 1626-67 have also been studied extensively since its discovery in 1977. 
Since then, 4U 1626-67 has transitioned through steady episodes of both spin-down and spin-up phases and and is currently in a spin-up period. 
During its first spin-up period between 1977 and 1990 the source's luminosity was estimated at $\approx10^{37}$\,erg/sec
and its pulse profile displayed a characteristic double peaked shape \citep{1983ApJ...270..711W, 2014MNRAS.439.1940B}.
As the  source moved closer to its first observed torque reversal, its pulse profile shape started to gradually lose its double peaked shape \citep{2014MNRAS.439.1940B}.
After the first torque reversal in 1990 the source entered a spin-down period, its flux decreased \citep{1997ApJ...474..414C} and the pulse profile changed to 
a broader shape that did not display the previously observed distinct peaks \citep{2007ApJ...660..605K}. In 2008, 4U 1626-67 underwent a new torque reversal and entered 
a new spin-up phase that is ongoing until today \citep{2010MNRAS.403..920J, 2012A&A...546A..40C}. During the second spin-up period the source luminosity increased by $\sim$2-3 times 
and the pulse profile shape returned to the double peaked shape \citep{2010MNRAS.403..920J, 2012A&A...546A..40C,  2014MNRAS.439.1940B}. 

The pulse profiles of accreting X-ray pulsars exhibit a variety of shapes, ranging from simple sinusoidal-like profiles,
to clear double-peaked shapes, and to more complicated broadened profiles with multiple peaks. The observed  variety of pulses among different -- or sometimes  same sources, reflects a variety of  possible emission patterns of the polar region of an accreting neutron star.
Depending on the mass accretion rate ($\sim$ source luminosity), the emission diagram of the accretion column may switch from a pencil-beam to a fan-beam pattern \citep{1975A&A....42..311B, 1976MNRAS.175..395B}.
At low luminosities, below  $\approx10^{37}$\,erg/sec,  the high anisotropy of the photon-electron scattering cross-sections in a high magnetic field of the order of $ 10^{12}$\,G \citep{1971PhRvD...3.2303C, 1974ApJ...190..141L}, leads to formation of the pencil beam pattern of  radiation, oriented parallel to the accretion column  \citep{1975A&A....42..311B}. 
However, at high mass accretion rates,  corresponding to luminosities above the critical value of  ${\rm L_c}\sim10^{37}$\,erg/sec \citep{1976MNRAS.175..395B, 1981A&A....93..255W},  a radiation dominated  shock is formed at the distance of  a $\sim$few km above the neutron star surface.  At high luminosities, the accretion funnel is filled with high density plasma slowly sinking in the gravitational field of the neutron star, resulting in increased opacity in the direction along the magnetic field axis. Consequently, the emerging X-ray photons predominantly   escape from the -- optically thin -- sides of the accretion funnel and the fan beam pattern of radiation is formed (Fig.~\ref{fig:sketch}).
Pencil-beamed emission is usually associated with single-pulse profiles 
or more complex shapes if one includes gravitational effects, and different obscuration mechanisms (e.g. \citeauthor{1992hrfm.book.....M} 1992). Double
peaked profiles are indicative of an emission pattern that is fan-beam dominated 
\citep[e.g.][]{1981ApJ...251..288N, 1983ApJ...270..711W, 1996BASI...24..729P, 1997A&A...319..507P, 2004A&A...421..235R}.

In the present paper we investigate the X-ray spectrum and pulse profile of 4U 1626-67 during two different luminosity states.
We present results of spectroscopic and timing analysis of an {\it XMM-Newton} observation performed in 2003 -- during the spin-down low luminosity period,
and the latest simultaneous {\it Chandra} and {\it RXTE} observations performed in 2010, during the current high luminosity, spin-up period. 
In Section \ref{sec:data} we describe  details of our data extraction along with their spectral and timing analysis, followed by 
interpretation of our results and discussion in Section \ref{sec:discussion} and conclusions in Section \ref{sec:conclusion}.

\section{Observations, data analysis and results}
\label{sec:data}

4U 1626-67 has been observed multiple times by all major X-ray telescopes. For the present work
we focus on the {\it XMM-Newton} observation performed in August 2003, the latest {\it Chandra} HETGs observation
performed in January 2010 and the {\it RXTE} observation that was performed simultaneously with the {\it Chandra} observation.
The details of the observations used in our analysis are listed in table~\ref{tab:obs2}.
The  temporal resolution of the EPIC pn instrument aboard {\it XMM-Newton} was sufficient  for  timing analysis of the 2003 data. For the 2010 data,  simultaneous {\it RXTE} observation allowed for high quality timing analysis of the source's light curve during the {\it Chandra} grating observation. Spectral analysis
was carried out using the {\small {XSPEC}} spectral fitting package, version 12.8.2 \citep{1996ASPC..101...17A}. Timing analysis
was performed using the standard tools of the {\small {XRONOS}} timing analysis software package, version 5.22.

\begin{table}
\centering
 \caption{Details of observations of 4U 1626--67 analyzed in  this paper}
 \label{tab:obs2}
 \begin{tabular}{@{}lcccc}
 Instrument & obsID & Date &  Duration$^{1}$ (ks)\\% &Count rate$^{2}$ \\
 \hline
    {\it XMM-Newton} & 0152620101&  2003-08-20  & 58\\% & $23.50\pm0.02$\\ 
  {\it Chandra}    & 11058 &2010-01-14  & 77 & \\%$2.26 \pm0.01$\\ 
  {\it RXTE}        & P95338-05-01-00 &2010-01-14  & 10\\% & $124.8\pm0.15$\\ 

  \hline
\end{tabular}

 \medskip
{$^{1}$Duration of filtered observations.\\}
%  $^{2}$Full bandpass.\\}
\end{table}

\subsection{XMM-Newton 2003 observation}
During the {\it XMM-Newton} observation MOS1 detector was operating in timing mode, while MOS2 and pn detectors were operating in imaging mode.
The MOS2 detector showed evidence of pile up. For this reason and since the effective area of pn at $\approx 7$\,keV is approximately five times higher, we only use the pn data for the present analysis.

\subsubsection{Spectral extraction and analysis}
We extracted the source spectrum from a 30$''$ circle centered at the source. Background was extracted in compliance with the 
latest EPIC calibration notes\footnote{http://xmm2.esac.esa.int/docs/documents/CAL-TN-0018.pdf} from a source-free region at the same {\small{RAWY}} position as the source region.
Spectral extraction was performed using the standard tools provided by the {\it XMM-Newton} Data Analysis software SAS, version 13.5.0.
The resulting spectrum was re-binned to ensure a minimum of 25 counts per energy channel.
4U 1626-67 is known for displaying strong emission features in the 0.5-1.5\,keV range.
The study of the low energy part of the spectrum of this source is beyond the scope of this paper
and has already been performed by \cite{2001ApJ...563..941S} and \cite{2007ApJ...660..605K}.
Therefore, in order to simplify our analysis, we ignored energy channels below 1.5\,keV. 
The remaining channels are sufficient to constrain the source continuum around  the iron line.

The spectral continuum was fit with an absorbed black body plus power law model.
The temperature of the black body was 0.30$\pm0.01$\,keV and the power law photon index 0.70 $\pm0.01$.
Our fit also required an exponential cutoff at  7.91$_{-0.23}^{+0.30}$\,keV with a folding energy of 29.2$_{-8.4}^{+8.7}$\,keV.
The spectral shape at this energy resembles a mild break in the power law, rather than a genuine exponential cutoff. Addition of the exponential cutoff model improves our fit by a $\Delta\chi^2$ of 30 for 2 dof.
Best fit parameters are presented in table~\ref{tab:cont2}
and are in agreement with the findings of previous authors \citep[e.g.][]{1995ApJ...449L..41A, 1998ApJ...500L.163O, 2007ApJ...660..605K} for the spin-down era of 4U 1626-67. 
We do not detect the iron line, placing an upper limit of $2.4$\,eV at 90\% confidence, for the equivalent width (EW) of a Gaussian emission line centered at 6.4\,keV with a
36\,eV width. The choice of the values for the line centroid and width was motivated by the corresponding best fit values   in the 2010 data, which are described in
section~\ref{Chandra}. 
The 1.5-10\,keV luminosity, calculated from the {\it XMM-Newton} fit, is $\approx1.3\times10^{36}$\,erg/sec, assuming a distance of 9\,kpc \citep{1998ApJ...492..342C}. 
Luminosity in the 0.01-100\,keV, extrapolated from the best fit model, is $\approx8.4\times10^{36}$\,erg/sec.
The data-to-model ratio vs energy is presented in Fig.~ \ref{fig:lines2}.

\subsubsection{Light curve extraction and analysis}
For our timing analysis we extracted a 2-12\,keV light curve from the pn data using standard SAS tools. Photon arrival times were corrected to the solar system barycentre. Using the tools provided in the {\small XRONOS} package version 5.22, 
we determined the pulse period during the {\it XMM-Newton} observation and created the pulse profile of the source in the 2-12\,keV range. The pulse period is measured at $\approx$ 7.67547(2)\,sec  
and the corresponding pulse profile is presented in Fig.~ \ref{fig:prof}.

\subsection{Chandra and RXTE 2010 observation}
 \label{Chandra}
The January 2010 {\it Chandra} observation of 4U 1627-67 was performed with the high energy grating. During the {\it Chandra} observation a simultaneous
10\,ks observation was also performed by {\it RXTE}. For our spectral analysis we used the data obtained by both {\it Chandra} and {\it RXTE}.
The {\it RXTE} data were also used for timing analysis.

\begin{figure}
\centering
\includegraphics[trim=0cm 1.20cm 0cm 0cm, clip=true, width=0.5\textwidth, angle=-90]{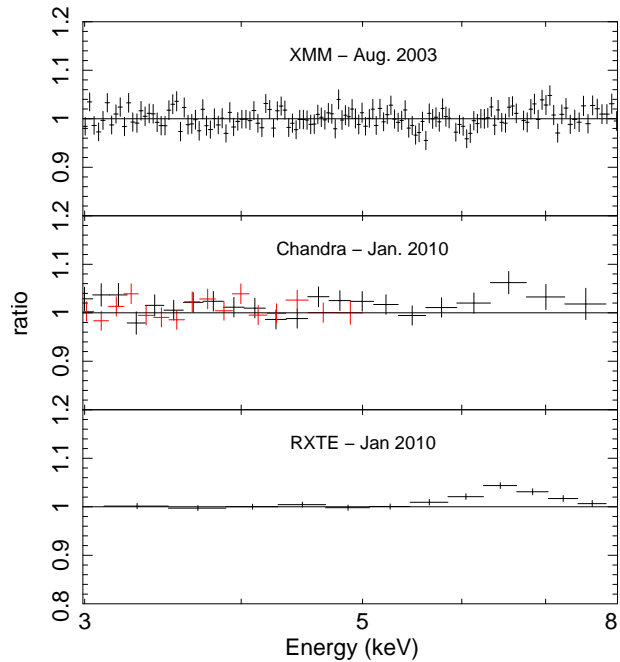}
\caption[Ratio of the data to the continuum model  for 4U 1626-67]{Ratio of the data to the continuum model for the 2003 XMM and 2010 Chandra and RXTE observations. The data have been rebinned for clarity; 
the 3-8\,keV energy range is shown. }
\label{fig:lines2}
\end{figure}

\subsubsection{Spectral extraction and analysis}
We extracted the spectra of both the medium energy grating (MEG) and the high energy grating (HEG). 
Extraction was executed using the standard tools\footnote{http://cxc.harvard.edu/ciao/threads/pointlike/} provided by the latest CIAO software (vers. 4.6.1).
In order to remain consistent with the {\it XMM-Newton} data analysis, we chose to ignore energy channels bellow 1.5\,keV in the {\it Chandra} data as well. 
We also ignored MEG energy channels above 5\,keV.
The {\it Chandra} data were not regrouped and were fitted using the standard $\chi^2$ method and the weighting technique suggested by \cite{1996ApJ...471..673C}.
We also extracted the source spectrum from {\it RXTE}-PCA standard-2 data. Spectral extraction and background subtraction were performed using the 
standard routines provided by the {\small FTOOLS} package, following the guidelines described in the {\it RXTE} Guest Observer Facility\footnote{http://heasarc.gsfc.nasa.gov/docs/xte/recipes/pca\_spectra.html}. 

Our analysis of the 2010 {\it Chandra} spectrum shows that since the 2003 {\it XMM-Newton} observation the photon index of the power law component 
has increased to a value of 1.13$\pm0.03$ and the black body temperature to 0.43$\pm0.01$\,keV. 
Most importantly, our analysis of the {\it Chandra} spectrum reveals the presence of an iron ${\rm K\alpha}$ line that was not present in the {\it XMM-Newton} observation of 2003. 
The line -- modeled using a simple Gaussian -- is located at 6.39$_{-0.01}^{+0.02}$\,keV, has a width of  $\sigma=36.4_{-11.3}^{+15.3}$\,eV and an EW of 18$_{-5.6}^{+6.2}$\,eV. 
Addition of the Gaussian line improves our fit by a $\delta\chi^2$ of 13 for 3 dof, giving the detection of the feature a 2.8\,$\sigma$ significance. 

The presence of the line
is confirmed by our fit of the 3-20\,keV, {\it RXTE}-PCA data, which strongly required a line with the best fit centroid energy of
6.64$_{-0.25}^{+0.20}$\,keV and the width of $\sigma=40^{+360}_{-40}$ eV. The EW of the line in the
{\it RXTE} fit, has a value of 35.1$_{-16.9}^{+17.4}$\,eV and is consistent -- within 1\,$\sigma$ errors -- with the {\it Chandra} value. 
While {\it RXTE} lacks the spectral resolution to constrain the line with the accuracy of {\it Chandra}, its larger effective area yields
a spectrum with substantially higher signal-to-noise ratio, thus providing a detection of the line with more than 4\,$\sigma$ significance ($\delta\chi^2=38.1$ for 3 dof).
The data-to-model ratios for the {\it Chandra} and {\it RXTE} observations are presented in Fig.~\ref{fig:lines2}. The plots are for the best-fit model 
of the continuum, without the Gaussian emission line. The presence of the iron emission line can be seen clearly, particularly in the {\it RXTE} data that have higher signal-to-noise ratio.
The 1.5-10\,keV luminosity, calculated from the {\it Chandra} fit, is $\approx4\times10^{36}$\,erg/sec. Luminosity in the 0.01-100\,keV, extrapolated from
the best fit model, is $\approx3\times10^{37}$\,erg/sec.
Best fit values for the spectral parameters obtained from {\it Chandra} and {\it RXTE} spectral analysis are presented in table~\ref{tab:cont2}.

\subsubsection{Light curve extraction and analysis}
\begin{figure}
\centering
\includegraphics[trim=0cm 1.20cm 0cm 0cm, clip=true, width=0.5\textwidth, angle=-90]{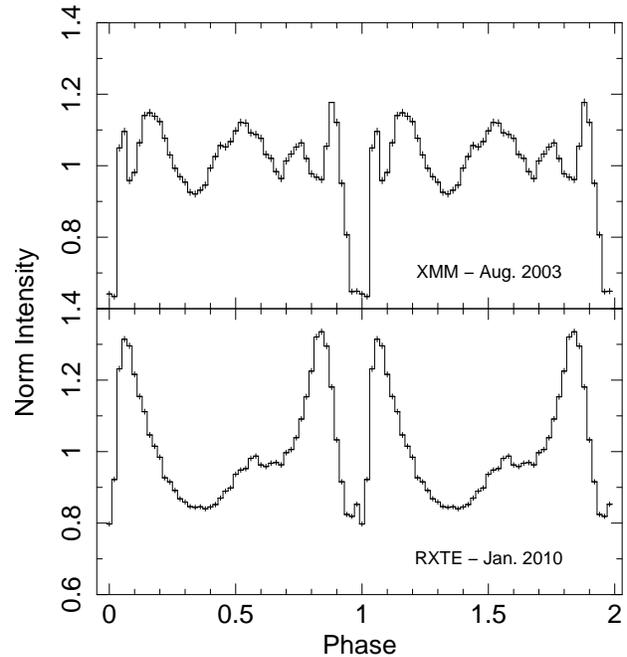}
\caption[Pulse profile of 4U 1626-67]{The 2-12\,keV pulse profile of 4U 1626-67 from  XMM and RXTE observations.
The switch to a distinct double horned shape in the 2010 RXTE observation is evident.
 The profiles have been arbitrarily shifted in  phase so that the minimum appears at the pulse phase 1.0.}
\label{fig:prof}
\end{figure}

For the timing analysis of the 2010 observation we made use of the {\it RXTE}-PCA good xenon data that have a time resolution of 1\,$\mu$s. 
Using the standard tools described in the previous sections, we extracted and analyzed the source's light curve. We measured the pulsation period
at $\approx$ 7.67797(6)\,sec and produced the corresponding pulse  profile in the 2-12\,keV range. The pulse profile is presented in Fig.~ \ref{fig:prof}.

\begin{table*}
 \centering
 \begin{minipage}{108mm}
\caption[Best fit parameters]{Best fit parameters for the XMM-Newton, Chandra and RXTE/PCA spectra. The errors are 1$\rm \sigma$.}
\label{tab:cont2}
  \begin{tabular}  {lcccccc}
  \hline
Model parameter   &  XMM-Newton 2003 & Chandra 2010 & RXTE-PCA 2010\\
 \cline{1-4}
 \hline

      nH$^{\it a}$  ($10^{21}\,{\rm cm^{-2}}$)                &    1.00       &1.00& 1.00\\
\hline

{\em Power Law}\\
  \hline
      $\Gamma$   				   &     0.70$^{\it b}$ $\pm0.01$       &1.13$\pm0.03$&1.04 $\pm0.02$\\
      
      norm$^{\it c}$                                   &     5.81$\pm0.07$      &34.9$\pm1.5$ &14.7$_{-0.6}^{+0.5}$\\
          
\hline
{\em Black Body}\\
  \hline
      kT (keV)   				   &     0.30$\pm0.01$       &0.43$\pm0.01$&0.61$\pm0.07$\\
      norm$^{\it d}$                                     &     1.19$\pm0.04$     &6.95$\pm0.41$&2.56$_{-0.38}^{+0.59}$\\
      \hline
{\em Iron Line}\\
  \hline
     Centroid E (keV)  				   &     6.39$^{\it e}$       &6.39$_{-0.01}^{+0.02}$ &6.64$_{-0.25}^{+0.20}$\\
     Width $\sigma$(eV)                                    &     36.4$^{\it e}$       & $36.4_{-11.3}^{+15.3}$&$<$400\\
     Flux$^{\it f}$                                &     $<$0.39       &7.66$_{-0.24}^{+0.26}$   &7.53$_{-3.64}^{+3.73}$\\
     EW (eV)  					   &     $<$2.41         &18.0$_{-5.6}^{+6.2}$&35.1$_{-16.9}^{+17.4}$\\
\hline
     $L_{1.5-10\,{\rm keV}}^{\it g}$        	   &     1.29         &4.00 &2.02$^{\it h}$\\
\hline
$\chi^2/{\rm dof}$                            &     1731/1693         &3073/3008 &38.15/38 \\
\end{tabular}
 \medskip

{$^{\it a}$ Parameter frozen at total galactic H\,I column density provided by the HEASARC nH tool \citep{1990ARA&A..28..215D, 2005A&A...440..775K}.\\
$^b$ With a high energy cut off at 7.91$_{-0.23}^{+0.30}$\,keV with an e-folding energy of 29.2$_{-8.36}^{+8.70}$\,keV, modeled using {\small {XSPEC}} model \texttt{highecut}. \\
% $^c$ With a high energy cut off at 7.33$\pm0.45$\,keV with an e-folding energy of 33.7$_{-7.47}^{+9.60}$\,keV.\\ 
 $^{\it c}$ $10^{-3}\,{\rm ph\,keV^{-1}\,cm^{-2}\,s^{-1}}$.\\
 $^{\it d}$ $10^{-4}\,{\rm L_{39}/D^{2}_{10kpc}}$ were ${\rm L_{39}}$ is luminosity in units of $10^{39}$ erg/sec and ${\rm D_{10kpc}}$ is distance in units of 10\,kpc. \\
 $^{\it e}$ Parameter frozen.\\
 $^{\it f}$  ${\rm 10^{-5}\,ph\,cm^{-2}s^{-1}}$.\\
 $^{\it g}$ $10^{36}$\,erg/sec.\\
 $^{\it h}$ Calculated in the 3-20\,keV range.}

\end{minipage}
\end{table*}

\section{Discussion}
\label{sec:discussion}

We have performed spectral and timing analysis of two different observations of 4U 1626-67 taken during low and high luminosity periods of the source. 
We detected a faint, narrow ${\rm K\alpha}$ emission line of iron at 6.4\,keV in the 2010 {\it Chandra} and {\it RXTE} data,  during the high luminosity state of the source. 
This feature was not present in 2003 (Fig.~\ref{fig:lines2}) when the source luminosity was less than $10^{37}$\,erg/sec. 
Furthermore, the appearance of the line coincides with a major alteration in the source's pulse profile. 
During the 2010 observation -- in which the iron line is detected -- the pulse profile of the source has 
a characteristic double peaked shape that is radically different from the pulse profile of the 2003 observation.
Below, we discuss the details of the iron emission line  and propose that  its appearance  is caused by the modification of the emission diagram of the accretion column caused by the change of the mass accretion rate. The same modification of the emission diagram leads to changes  in the shape of the pulse profile. 

\subsection{The iron ${\rm K\alpha}$ line}

Parameters of the iron lines  observed in the spectra of accreting X-ray pulsars can vary significantly from source to source.  
Depending on the environment and the ionization state of iron, the energy of the  K-shell  line  can vary from 
$\approx6.4$\,keV in the case of fluorescent line of neutral and weakly ionized iron to  $\approx6.9$\,keV in the case of the resonant line of  highly ionized, hydrogen-like Fe{\small XXVI}. The observed line width also varies in the broad range  from a few tens of eV to $\ga 0.5$\,keV. Typical values of the equivalent width of the line are of the order of 0.1 to 0.4\,keV \citep[e.g.][]{1983ApJ...270..711W, 1995A&AS..109....9G, 2010ApJ...715..947T}. From a theoretical point of view, the iron line  in X-ray pulsars may appear due to reflection of the primary X-ray emission  off the accretion disk and/or the donor star \citep[e.g.][]{1974A&A....31..249B}. It can also be due to Alfv{\'e}n shell emission, as proposed by  \cite{1980A&A....87..330B}.
In their work, \citeauthor{1980A&A....87..330B} argued that the iron emission line originating in the moderately optically thick, highly ionised plasma of the  Alfv{\'e}n shell are expected to be centered at $\ga 6.5$\,keV, are broader and brighter, with equivalent widths exceeding $\ga 0.3$\,keV. On the contrary, fluorescent  iron  line  produced in highly magnetised X-ray pulsars  by reflection from the accretion disk and the donor star surface, are expected to be generally narrower, centered at $\approx$6.4\,keV and have a moderate equivalent width  of the order of 50-100\,eV.

 The iron ${\rm K\alpha}$ line detected in the spectrum of  4U 1626-67 during the 2010 {\it Chandra} observation is centered at 6.4\,keV, has a small width of $\approx$36\,eV and an the equivalent width of $\approx$18\,eV. 
These line properties  are more consistent with reflection from either the cool surface of the donor star or from the accretion disk. Distinguishing between the two latter possibilities, we note the following. 
For a binary system consisting of a $\rm 1.4 M_{\odot}$ NS accretor and a $\rm 0.02 M_{\odot}$ WD donor \citep[e.g.][]{1998ApJ...492..342C}, the Roche lobe of the WD subtends a solid angle of $\rm \sim 5.1\times10^{-2}\,sr$ as viewed from the neutron star. Assuming that the emission source is located in the accretion column, at $\sim$15\,km above the the disk plane and that the disk is truncated at the magnetospheric radius (see eq.~\ref{eq:mag} and discussion below), the disk will subtend a solid angle of $\rm \sim 1.3\times10^{-2}\,sr$ as viewed from the emission source, ignoring possible flaring of the disk. This is smaller than the solid angle subtended by the WD donor, however, disk
flaring and non-isotropic emission pattern can change this number and the relative
contributions of the disk and the surface of the donor star. Indeed, the width of
the line measured by the high energy grating, $\approx 36$ eV, corresponds to velocities of $\approx 1700$ km/s. Such velocities are more typical for  the accretion disk, rather than for the surface of the donor star, suggesting that the majority of the line emission originates from the disk. 
Furthermore, actual contributions of the accretion disk and the donor star to the observed fluorescent line flux are determined by not only the solid angle, but also depend on the emission diagram of the primary emission and the angles of the line of sight to the normal of the surfaces of the disk and the star.  The latter are mainly defined by the inclination of the binary system. As for the former, in the case of  the fan beam, for example,  it is quite possible that much larger  fraction of the accretion column emission is intercepted by the accretion disk, rather than by the donor star. 

The magnetospheric radius of the neutron star can be estimated as follows \citep[e.g.][]{1977ApJ...217..578G}.
\begin{equation}
 R_{m} = {\left( {\frac{{B^2}{R_{NS}}^6}{{\dot M}\sqrt{2GM_{NS}}} } \right)^{2/7}}
 \label{eq:mag}
\end{equation}
The magnetic field of the  neutron star in  4U1626--67 is known from the cyclotron line measurements,  $B \approx 3\times10^{12}$ \citep[e.g.][]{1998ApJ...500L.163O}. Assuming a $M_{NS}=1.4{M_ \odot }$ neutron star of the radius of $R_{NS}=12$ km,  accreting 
at the mass accretion rate of ${\dot M}=\,2.7\times10^{-9}\,{M_ \odot }/{\rm yr}$ corresponding to the observed luminosity of $\approx3\times10^{37}$\,erg/sec, the magnetospheric radius is $R_{m}\approx6.8\times10^{8}$\,cm.
This value is very close to the pulsar corotational radius $R_{co}=(G\,M_{NS}\,P_{\rm spin}^2/4\,\pi^2)^{1/3}\approx6.5\times10^{8}\,cm$. Hence, the pulsar is expected to be near spin equilibrium 
($ R_{m}\approx R_{co}$), \citep[e.g.][]{1975SvAL....1..223S, 1976SvAL....2..133L, 1985AcA....35..185Z}. The accretion disk will be disrupted by the magnetic field close to the magnetospheric radius, i.e. to the first approximation, the inner disk radius can be estimated as $R_{\rm in}\approx 6.8\times10^{8}$\,cm \citep[e.g.][]{1972A&A....21....1P,2012MNRAS.421...63R}.
The outer disk radius will extend to a substantial fraction \citep[up to 0.8, e.g.][]{2001NewAR..45..449L} of the Roche lobe radius, which is $\approx2.2\times10^{10}$\,cm, assuming a white dwarf donor
of 0.02\,${M_ \odot }$ \citep[e.g.][]{1998ApJ...492..342C}.

The line is resolved by the Chandra HETG at the 98\% confidence interval.  The 90\% error interval for the line width, $25.1-51.7$\,eV, corresponds to the range of line of sight velocities of $\approx 1178-2426$\,km/sec. 
For the disk inclination angle of $20\degr$ \citep[e.g.][]{1990A&A...234..195V, 1998ApJ...492..342C},  this corresponds to the  3D velocity range of $3443-7092$\,km/sec.
Keplerian velocity reaches these values at  the distance of   $(3.7-15.7 )\times 10^8$\,cm from the compact object, with the best fit value of the line centroid -- $36.4$ keV, corresponding to the distance of $7.5\times 10^8$\,cm.
These numbers are consistent with our estimate of  the magnetospheric radius ($R_{m}\approx6.8\times10^{8}$\,cm, see above), suggesting that the fluorescent  line is produced  near the magnetospheric boundary of the neutron star, where the Keplerian accretion disk is expected to be truncated by the neutron star magnetic field.
At these radii, an accretion  disk that is heated through viscus dissipation \citep{1973A&A....24..337S}, cannot reach a temperature
higher than $\approx5-6\times$ $10^4$\,K.
Obviously, at these temperatures any iron that is present in the disk will be in a low ionization or neutral state and its fluorescent K-alpha line will be centered at 
$\approx$6.4\,keV \citep[e.g.][]{1967RvMP...39...78B} as observed.

\begin{figure}
\centering
\includegraphics[trim=1.2cm 1.2cm 1.2cm 1.2cm, clip=true, width=0.47\textwidth, angle=0]{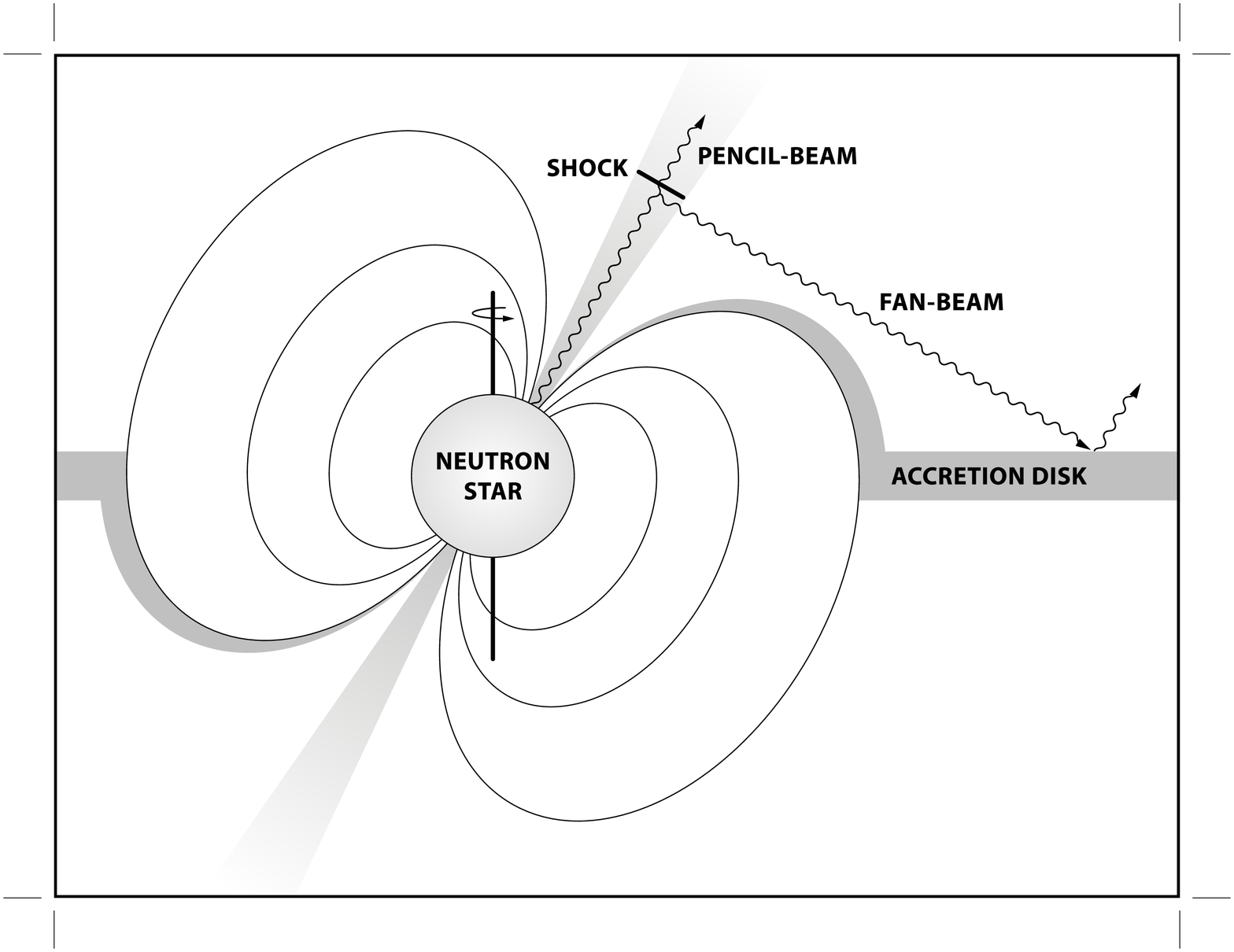}
\caption[Schematic representation of X-ray pulsar emission]{Schematic representation of the pencil and fan-beam emission in an accreting X-ray pulsar. Note that the drawing is only aimed to illustrate the difference of the disk illumination at different emission patterns of the accretion column and is not meant to realistically reproduce the geometry  of  an accreting strongly magnetised neutron star. In particular, for viewing clarity, the distance of the shock from the NS surface has been exaggerated, as well as the size of the neutron star itself, whereas  the inner disk radius has been depreciated.}
\label{fig:sketch}
\end{figure}

Although the observed line energy and width, are consistent with what should be expected for this system, the line itself is quite faint. 
With an EW of $\approx$18\,eV, it is significantly fainter than the EWs of $\approx$ 40-100\,eV  
of iron lines usually observed in disk reflection spectra of non-pulsar LMXBs \citep[e.g.][]{ 2010ApJ...720..205C}.
This could be due to the fact that the truncated accretion disk around the highly magnetised neutron star subtends a significantly smaller solid angle than in the case of  "normal" LMXBs.
However, the  non-isotropic emission diagram of the accretion column can compensate for the smaller solid angle, for example in the case of the fan-beam as discussed above (Fig.~\ref{fig:sketch}). Indeed, narrow fluorescent lines of iron, of appreciable equivalent width in the $\sim 70$ eV range, were previously detected in the spectra of some accreting X-ray pulsars in low-mass X-ray binary systems. For example, a narrow  line at $\approx$6.4\,keV was  detected   in GX 1+4    \citep[ EW$\sim$70\,eV, ][]{2005ApJ...627..915P} and in  Her X-1    \citep[EW$\sim$65\,eV][]{2000PASJ...52..223E, 2003A&A...401..265N}. 
The 6.4 keV line, we found in 4U 1626-67, is significantly  fainter than in those sources.

If the reflected component is not strongly reduced due to the geometry of the system (the detection of O and Ne emission features by \cite{2007ApJ...660..605K} and \cite{2001ApJ...563..941S} suggests it is not), then the faintness of the iron line can be explained by the fact that 4U 1626-67 is an UCXB 
with a C/O-rich donor \citep{2001ApJ...563..941S, 2002AJ....124.3348H, 2006MNRAS.370..255N, 2006A&A...450..725W}. 
In \cite{2013MNRAS.432.1264K} and \cite{2014MNRAS.442.2817K} we demonstrated that the iron ${\rm K\alpha}$ line
is strongly attenuated in reflection spectra of C/O or O/Ne dominated disks. This is due to screening of the presence of iron by the overabundant oxygen.
In particular, in the C/O dominated material the main interaction process for an $E\geq7.1$ keV photon is  absorption by oxygen rather than by iron, contrary to the
case of ``standard" LMXBs with main sequence or red giant donors where the accretion disk has a chemical composition close to Solar.

To further investigate this hypothesis, we ran our simulation from  \citet{2013MNRAS.432.1264K}  for a primary radiation with a power law  spectrum with a photon index of 1.1. For simplicity, we assumed that the primary radiation is emitted isotropically by a point source above the disk surface, in a lamppost configuration. We have collected the reflected emission for  viewing angles in the range  ${\rm 15\degr-25\degr}$.
We found that in order to produce an output spectrum with the iron ${\rm K\alpha}$ line with an EW of $\approx18$\,eV -- as observed -- we need an O/Fe ratio in the disk that is 68 times the solar value. Note that the maximum value of the O/Fe ratio that corresponds to the  chemical composition of a C/O white dwarf -- in which all hydrogen and helium has been converted to carbon and oxygen -- is $\approx77$ times the solar value. This result suggests that the donor star in  4U1626--67 is a C/O or O/Ne/Mg white dwarf, in a perfect agreement with the  previous work which proposed  that this source is a UCXB with a C/O white dwarf donor \citep{2001ApJ...563..941S, 2007ApJ...660..605K}.  To conclude, we note that a  lamppost  geometry is
obviously not an accurate representation of the emission of the accretion column.  However, it can serve as a sufficient first approximation which demonstrates that the small equivalent width of the iron line  in the spectrum of 4U 1626-67 can be easily explained in terms of  reflection from a C/O rich disk.

\subsection{The origin of the iron line variability.}
\label{appearance}

The iron line was not detected in the 2003 {\it XMM-Newton} data,  with the  upper limit of  $\approx 2.4$ eV ($1\sigma$) on its  equivalent width.  A possible explanation for the line variability  is suggested by the timing analysis of the source.
In particular, we notice that the line is present during the high luminosity state of the source  in which  the pulse profile has switched to a characteristic double peaked shape.

We argue that both the appearance of the fluorescent  line of iron  and the change of the shape of the pulse profile are the result of a major modification in the emission diagram of the accretion column.  Such a modification may be caused by the increase of the  mass accretion rate, as proposed by  \citet{1975A&A....42..311B}. Observationally, in the case of 4U1626--67, the increase of the mass accretion rate manifests  itself as a more than 3-fold increase of the source luminosity and acceleration of the neutron star spin.

At low luminosity, the emission of the accretion column is concentrated in a beam that is oriented along the magnetic field axis (pencil-beam emission) \cite{1975A&A....42..311B}. Depending on the angle between the rotation axis and magnetic dipole of the neutron star, the magnetic field axis -- and hence the beamed radiation  -- will be mostly directed away from the accretion disk  (Fig.~\ref{fig:sketch}).  Due to the decreased flux towards the disk, disk reflection features in the observed spectrum will be significantly suppressed, including the iron line emission. Note that the latter is  weakened  further due to the C/O dominated chemical composition of the accretion disk.

As mass accretion rate increases, the emission diagram of the accretion column changes to the fan-beam pattern \cite{1976MNRAS.175..395B}. This modification in the emission diagram of the accretion column, causes most of the radiation to be beamed towards the accretion disk (Fig.~\ref{fig:sketch}), resulting in a significant boost of the emission that is reflected off the disk and the appearance of the detectable ${\rm K\alpha}$  line of iron. As the disk is truncated by the magnetic field of the neutron star at large distance form the latter, relativistic broadening of the line is insignificant and the line is rather narrow.  The line is also rather faint, due to the fact that the disk is made of  C/O-enriched material. The fan-beam diagram  of the emission from the accretion column further manifests itself  throughout the  characteristic double peaked shape of the pulse profile. 

In this picture, we can predict that the iron line should  have been suppressed during the entire spin-down period from $\sim 1990$ to $\sim 2008$. In order to check this prediction, we analysed two earlier Chandra observations performed in 2000 (with duration of 39 ksec) and 2003 (95 ksec), during the spin-down era. We detect no emission lines in the 6-7\,keV energy range in either of the these observations. We obtained an upper limit (1\,$\sigma$) on the equivalent width of the 6.4 keV iron line of  24\,eV  in the 2000 data and a 7.1\,eV  in the 2003 data. In both cases the  {\it Chandra} spectra were fitted with the same spectral model as  {\it XMM-Newton} data, and produced  similar values of best fit parameters. The 6.4 keV iron line was not detected in the 1994 {\it ASCA} data  neither, although with a fairly unconstraining upper limit of  of 33\,eV \citep{1995ApJ...449L..41A}.  In all three observations the source  was in the same low luminosity regime. These non-detections of the line provide further support to the picture outline above.

\section{Summary and conclusions}
\label{sec:conclusion}

We have discovered the appearance of a narrow Fe ${\rm K\alpha}$ emission line in the spectrum of 
4U 1626-67 correlated with a major change in the shape of its pulse profile.  The line was detected at a moderate level of $EW=18_{-5.6}^{+6.2}$ eV during a period of high luminosity in 2010, when the pulse profile of the source had a characteristic double peaked shape. The line was not detected in 2003 with an upper limit of 2.4 eV ($1\sigma$) when the source was in a low luminosity state  and the shape of its pulse profile had a clearly different, broader shape. We argue that both changes are caused by the same reason -- a significant
modification of the emission diagram of the accretion column, from a pencil beam to a fan beam pattern. This change was caused by an increase in the mass accretion rate, as was theoretically predicted by Basko \& Sunyaev in \citeyear{1976MNRAS.175..395B}. The lower than typical equivalent width of the iron line in the high luminosity state is due to C/O dominated chemical composition of the accretion disk.

\section*{Acknowledgements}
The authors would like to thank Deepto Chakrabarty and Georgios Vassilopoulos for stimulating discussion and valuable advice and Cheryl Woynarski for her artistic input. MG acknowledges hospitality of the Kazan Federal University (KFU) and  support by the Russian Government Program of Competitive Growth of KFU. The authors would like to thank the anonymous referee for useful and constructive comments which helped to improve the paper.

\bibliographystyle{mn2e}
\bibliography{General}

\label{lastpage}
\end{document}